\documentclass[conference]{IEEEtran}
\IEEEoverridecommandlockouts
\usepackage{cite}
\usepackage{amsmath,amssymb,amsfonts}
\usepackage{algorithmic}
\usepackage{graphicx}
\usepackage{textcomp,tikz,hyperref}
\usepackage{nccmath}
\usepackage{mathtools}
\usepackage{xcolor}
\def\BibTeX{{\rm B\kern-.05em{\sc i\kern-.025em b}\kern-.08em
    T\kern-.1667em\lower.7ex\hbox{E}\kern-.125emX}}
\usepackage{cuted}
\ifCLASSOPTIONcompsoc
    \usepackage[caption=false, font=normalsize, labelfont=sf, textfont=sf]{subfig}
\else
\usepackage[caption=false, font=footnotesize]{subfig}
\fi
\usepackage{algorithm}
\usepackage{algorithmic}

\makeatletter
\newcommand\fs@norules{\def\@fs@cfont{\bfseries}\let\@fs@capt\floatc@ruled
  \def\@fs@pre{}%
  \def\@fs@post{}%
  \def\@fs@mid{\kern3pt}%
  \let\@fs@iftopcapt\iftrue}
\makeatother
\floatstyle{norules}
\restylefloat{algorithm}

\DeclarePairedDelimiter\floor{\lfloor}{\rfloor}

\usepackage{braket}

\newcommand\copyrighttext{%
  \footnotesize \textcopyright 2021 IEEE. Personal use of this material is permitted. Permission from IEEE must be obtained for all other uses, in any current or future media, including reprinting/republishing this material for advertising or promotional purposes, creating new collective works, for resale or redistribution to servers or lists, or reuse of any copyrighted component of this work in other works. 
}
\newcommand\copyrightnotice{%
\begin{tikzpicture}[remember picture,overlay]
\node[anchor=south,yshift=10pt] at (current page.south) {\fbox{\parbox{\dimexpr\textwidth-\fboxsep-\fboxrule\relax}{\copyrighttext}}};
\end{tikzpicture}%
}

\begin{document}
\title{A Quantum-Inspired Classical Solver for Boolean $k$-Satisfiability Problems
}

\author{
\IEEEauthorblockN{S. Andrew Lanham, Brian R. La Cour}   
\IEEEauthorblockA{\textit{Applied Research Laboratories} \\
\textit{The University of Texas at Austin}\\
Austin, TX, 78713-8029 USA \\
andrew.lanham@arlut.utexas.edu, blacour@arlut.utexas.edu}

}
\maketitle
\copyrightnotice

\begin{abstract}
In this paper we detail a classical algorithmic approach to the $k$-satisfiability ($k$-SAT) problem that is inspired by the quantum amplitude amplification algorithm. This work falls under the emerging field of quantum-inspired classical algorithms. To propose our modification, we adopt an existing problem model for $k$-SAT known as Universal SAT (UniSAT), which casts the Boolean satisfiability problem as a non-convex global optimization over a real-valued space. The quantum-inspired modification to UniSAT is to apply a conditioning operation to the objective function that has the effect of ``amplifying” the function value at points corresponding to optimal solutions. We describe the algorithm  for achieving this amplification, termed ``AmplifySAT," which follows a familiar two-step process of applying an oracle-like operation followed by a reflection about the average. We then discuss opportunities for meaningfully leveraging this processing in a classical digital or analog computing setting, attempting to identify the strengths and limitations of AmplifySAT in the context of existing non-convex optimization strategies like simulated annealing and gradient descent. 
\end{abstract}



\section{Introduction}
The development of new quantum algorithms has given rise to rich tradeoffs between classical and quantum solutions for different problems. Quantum computing's promise lies in the fact that there are many problems for which quantum algorithms produce a significant speedup over the best known classical algorithms. Yet at the same time, careful scrutiny of the differences between quantum and classical algorithms has led to new classical analogues of quantum algorithms, sometimes promising the same computational advantages. The resulting area of research, termed quantum-inspired algorithms, means that our understanding of the delineation between quantum and classical algorithms is continually evolving.  Quantum-inspired design approaches have produced novel algorithms in a variety of application areas, perhaps most concretely in recommendation systems \cite{tang2019quantum}, but also in reinforcement learning \cite{dong2010robust}, constraint satisfaction \cite{barak2015beating}, and more general areas of machine learning and computational linear algebra \cite{chia2020sampling, narayanan1996quantum, chia2018quantum, chen2019quantum}. 

Recent quantum-inspired algorithms leverage similarities between quantum state preparation assumptions and input assumptions for classical algorithms \cite{tang2018quantum,chia2020sampling}. In these works, quantum state preparation assumptions are shown to be equivalent to a set of classical assumptions involving \textit{sampling access} to a vector or matrix. Sampling access refers to a situation in which a vector or matrix can be normalized and the entries treated as a discrete distribution to be sampled from. The similarity between sampling access and state preparation allows for a ``dequantization" of a variety of quantum algorithms by adopting the assumptions in a classical setting and proposing analogous classical processing. A thorough characterization of this framework with applications to quantum and classical machine learning algorithms is contained in \cite{chia2020sampling}. 

Other families of quantum algorithms are much less likely to admit a quantum-inspired classical analog because their speedup relies on uniquely quantum resources like entanglement. In this work, we focus on an algorithm inspired by amplitude amplification (AA). At a high level, AA uses principles of quantum coherence to evolve a quantum state. Coherence is not a uniquely quantum phenomenon, and, as we will explore in this work, there is potential to use it to provide speedups for classical solvers. Additionally, the focus on AA places this work in a different area than recent dequantized algorithms based on quantum state preparation assumptions. 
 
AA algorithms find application in areas including unstructured search \cite{Grover1997}, function optimization \cite{ahuja1999quantum}, and some types of numerical estimation \cite{abrams1999fast, grover1998framework}. More generally, AA can be applied to the output state of almost any quantum algorithm to boost the probability of success for that algorithm \cite{brassard2002quantum}. Each algorithm in the AA family works by increasing the amplitude of a desirable element in a quantum superposition by $\mathcal{O}(1/\sqrt{N})$ for each loop of the algorithm. It follows that with $\mathcal{O}(\sqrt{N})$ repetitions of this algorithm, the desired element is measured with probability $\mathcal{O}(1)$. In this sense, AA is a way of boosting the probability of measuring a given element in a set (typically large) of possibilities, and as such, is intimately related to sampling access to a vector.

In a classical setting, therefore, it is not immediately obvious how to introduce an algorithm that is designed for sampling access. It is rare that sampling access to classical data occurs naturally, and thus it must often be considered as a pre-processing assumption, as occurs in much quantum-inspired work \cite{tang2018quantum, chia2020sampling}. In contrast, classical data can often be assumed to admit efficient \textit{query access}, which simply refers to the fact that individual entries of a vector or matrix can be efficiently retrieved. Since this work details a classical analog to a quantum AA algorithm, we are forced to address the presence or lack of sampling access. In this work we make no assumption of sampling access because it would be far too costly and unnatural for the problem setting. This leads to an exploration of the interplay between query access and sampling access; in particular, we explore whether algorithms designed assuming sampling access can be meaningfully used in situations in which only query access is available. A quantum-inspired AA algorithm is a prime candidate for exploring this question. 

Boolean satisfiability is an important class of problem that can be treated as an unstructured or semi-structured search, and is therefore amenable to AA. Boolean satisfiability concerns questions of determining whether there exists a binary string $x \in \{0, 1\}^n$ for which a given Boolean function is satisfied, that is, evaluates to $1$. 
Considerable effort has been focused on developing better SAT solvers. For example, SAT solvers have been designed based on a variety of heuristic search algorithms including backtracking \cite{chaff, een2003extensible}, conflict-driven clause learning (an evolution of backtracking) \cite{marques1999grasp}, local search \cite{dueck1993new, brueggemann2004improved, seitz2005focused}, and simulated annealing \cite{spears1993simulated}. A classical randomized algorithm for $3$-SAT problems achieving average complexity within a polynomial factor of $\mathcal{O}(1.334^n)$ was proposed by Sch\"{o}ning \cite{schoning1999probabilistic}. These algorithms and similar ones have also led to a number of hardware-based SAT solvers achieving significant speedups \cite{yin2017efficient,davis2008designing, basford2016impact}. With quantum computers, Grover's algorithm stands out as the most relevant algorithm for solving SAT problems \cite{Grover1997}. Its generalization, AA, can be used to achieve quadratic speedups over the best known classical algorithms for search \cite{brassard2002quantum}.

A separate algorithmic paradigm for solving SAT problems casts discrete Boolean SAT problems as global minimizations over a bounded real-valued space \cite{gu1994global, johnson1989neural, kyrillidis2020fouriersat, gu1996algorithms}. With the conversion to a continuous domain, the resulting minimization task becomes amenable to a variety of solution strategies from optimization, such as gradient descent and its generalizations. Nevertheless, the objective function that results from this conversion is likely non-convex and so solution strategies remain largely heuristic in design. A SAT solution strategy broadly based on the problem formulation of \cite{gu1999optimizing}, known as UniSAT, was proposed in \cite{ercsey2011optimization} and an analog solver was presented in \cite{yin2017efficient}. The work of \cite{ercsey2011optimization} significantly extends the original formulation by proposing Lagrange multiplier methods as a strategy for avoiding local minima. A continuous-time dynamical system formulation is presented with efficient convergence to the global minimum, but at exponential energy cost. The work of \cite{yin2017efficient} follows this up with an analog circuit demonstration implementing this dynamical system. In this work, the discussion of quantum-inspired algorithms with possible applications to analog solvers loosely relates to the field of quantum emulation, which proposes classical analog circuits for executing quantum algorithms \cite{LaCour&Ott2015,LaCour&Ostrove2017}. 

This work continues development under the UniSAT problem model by proposing a modification of the objective functions originally presented in \cite{gu1999optimizing} and \cite{ercsey2011optimization}. Modifications to the objective function are common in optimization, for example, when introducing barrier functions to convert a constrained optimization to an unconstrained one. Our modification, termed AmplifySAT, is inspired by the quantum AA algorithm, and has the qualitative effect of ``amplifying", or ``deepening" the global optimum of an objective function encoding a SAT problem. Transformations of the objective function that target the extrema are not common, but have been explored in some works, e.g.,  \cite{karandashev2010binary, karandashev2014attraction}. The iterative nature of AA means that the degree of amplification of the extrema can be coarsely controlled. We continue in Section \ref{sec:intro_problem} with a review of the UniSAT problem formulation. In Section \ref{sec:mod_cost} we propose using AmplifySAT to modify the objective function. We conclude in Section \ref{sec:conclusion} with an exploration of the numerous possibilities to follow-up this investigation. In particular, we suggest reconsidering the performance of various standard non-convex optimization strategies with objective functions conditioned by AmplifySAT. In this sense, this work does not propose a brand new optimization algorithm, but simply views AmplifySAT as a conditioning stage that may improve the performance of existing optimization approaches. 


\section{The UniSAT Problem Model}
\label{sec:intro_problem}
We begin by reviewing the $k$-SAT problem and its UniSAT formulation. The goal when solving such a problem is to find the values of $n$ variables $\mathbf{x} \in \{0, 1\}^n$ that satisfy a propositional formula $F$. The formula is assumed to be in the conjunctive normal form, that is, a conjunction of clauses $C_m, m = 1, \ldots, M$ with 
\begin{equation}
    \label{eq:bool_sat_problem}
    F(\mathbf{x}) = \bigwedge_{m = 1}^{M} C_m 
\end{equation}
We have $F(\mathbf{x}) : \mathbb{F}_2^{n} \mapsto \{0, 1\}$, and the function returns $1 = \textrm{TRUE}$ if the function is satisfied for input $\mathbf{x}$, and otherwise it returns $0 = \textrm{FALSE}$. A clause is a disjunction between $k$ variables. Complements of variables, e.g. $\bar{x}_i$ for variable $x_i$ are valid variables as well. As an example, a $k$-SAT clause could take the form
\begin{equation}
    C_1 = (x_1 \lor x_4 \lor \bar{x}_7 ) \, . 
\end{equation}

The UniSAT problem model expresses a $k$-satisfiability problem as a function over a bounded continuous space so that it can be solved using different classes of solvers \cite{gu1999optimizing,ercsey2011optimization}. In the $k$-UniSAT relaxation, each clause $C_m$ is represented by a continuous function $K_m(\mathbf{s}) : [-1,1]^n \mapsto [0,1] $ given by 
\begin{equation}
    \label{eq:clause_fn}
    K_m(\mathbf{s}) = 2^{-k} \prod_{i = 1}^{n} (1 - c_{m, i}s_i)
\end{equation}
The objective function is defined over the $n$-dimensional hypercube, i.e. $[-1, 1]^n$. The vertices of this hypercube correspond to the binary inputs to the Boolean satisfiability problem of Equation (\ref{eq:bool_sat_problem}). This means that $\mathbf{s}$ obeys the equivalence that $s_i = -1 \mapsto \textrm{FALSE}$ and $s_i = +1 \mapsto \textrm{TRUE}$ for each entry $s_i$ of $\mathbf{s}$.  We express the set of vertices for an $n$-variable problem as
\begin{equation}
    \label{eq:vertices}
    \mathcal{V} = \{ \mathbf{x} : \mathbf{x} \in \mathbb{R}^n, x_i \in \{-1, 1\} \; \forall i \in [n] \} \, .
\end{equation}
The clause variables $c_{m,i}$ take value $1$ when variable $x_i$ appears in clause $m$, $-1$ if $\bar{x}_i$ appears, and $0$ if neither $x_i$ nor $\bar{x}_i$ appear in clause $m$. From the clause functions $K_m(\mathbf{s})$, \cite{gu1999optimizing} defines an objective function 
\begin{equation}
    \label{eq:cost_fun}
    E(\mathbf{s}) = \sum_{m = 1}^{M} K_m(\mathbf{s}) \, , 
\end{equation}
with the property that if $E(\mathbf{s}^*) = 0$ for $\mathbf{s}^* \in \mathcal{V}$ then $\mathbf{s}^*$ corresponds to a satisfying solution. The satisfying solutions, therefore, are encoded as vectors $\mathbf{s} \in \mathcal{S}^* \subseteq \mathcal{V}$ achieving the global minimum. Here we have defined $\mathcal{S}^*$ as the set of vectors mapping to satisfying solutions to a given satisfiability problem $F$. Clearly, the objective function $E(\mathbf{s})$ may also be evaluated at points in the hypercube that are not vertices to assist in finding the global minimum. 




\section{A Modification to the Objective Function Inspired by Amplitude Amplification}
\label{sec:mod_cost}
The role of AA in satisfiability is to boost the probability amplitudes associated with the satisfying solutions to a problem of interest. When considering a quantum-inspired version of this algorithm, we must contend with the numerous differences between classical and quantum processing. Fundamentally the difference can be understood in terms of how information encoded in a quantum or classical state is efficiently accessed. The types of access for each setting are best understood as ``sampling access" and ``query access," and their relevance has been highlighted in other quantum-inspired classical algorithms \cite{tang2018quantum,chia2020sampling}. We review these types of access now. The quantum paradigm typically admits efficient sampling access to the set of feasible solutions when they are encoded in a quantum state. This is because in a quantum state, the formalism of probability amplitudes and quantum measurements is integral to how data is accessed. Sampling access is efficiently realized through the process of quantum measurement. With measurement, each possible outcome is observed with frequency proportional to the modulus-squared of the probability amplitude of that outcome. In a classical setting there is no efficiently realizable analogy to this notion, in general. In contrast, only the output values corresponding to individual inputs can  be ``queried"  efficiently, meaning that an individual input must be specified and information (e.g. ``satisfying" or ``not satisfying") for only that input is returned.

The AmplifySAT algorithm presented in this section, being inspired by AA, is naturally optimized for sampling access. Since there is no well-defined notion of sampling from a classical objective function, we will have to address whether an algorithm designed for sampling access is even useful in a setting where only query access is efficiently available. Despite this apparent incompatibility, we will see that AA has observable effects on query accesses as well, and that it has the potential to improve the performance of algorithms relying only on query access. In particular, the amplification step introduced in AmplifySAT affects properties of the objective function relevant to the convergence of non-convex solvers that only access the function locally. The algorithm works as follows. First, we construct a new objective function $V(\mathbf{s})$ based on the original objective function $E(\mathbf{s})$, but for which the satisfying arguments $\mathbf{s} \in \mathcal{S}^*$ are maxima instead of minima. AmplifySAT then iteratively transforms $V(\mathbf{s})$ into a conditioned function $V_{\ell}(\mathbf{s})$ for the $\ell^{\textrm{th}}$ iterate, whose extremal points are ``amplified" using a two-step process similar to multiplication by an oracle operator followed by reflection about the average. 

\subsection{Modified Objective Function}
The objective function $E(\mathbf{s})$ of Equation (\ref{eq:cost_fun}) has a variety of useful properties; namely, it's constrained to the interval $[0, M]$ so it's non-negative. Furthermore, the set of binary inputs to the SAT problem can be mapped to the vertices $\mathcal{V}$ of the $n$-dimensional hypercube $[-1, 1]^n$. 
But in order to benefit from amplification effects, we must consider an objective function for which the optimal points are the global \textit{maxima}. This is easily obtained as a modification of $E(\mathbf{s})$ by substituting each clause function $K_m(\mathbf{s})$ by $1 - K_m(\mathbf{s})$, giving
\begin{equation}
    \label{eq:cost_max}
    V(\mathbf{s}) = \sum_{m = 1}^{M} (1 - K_m(\mathbf{s})) \, .
\end{equation}
Since each $K_m(\mathbf{s}) \in [0,1]$, each term $(1 - K_m(\mathbf{s}))$ in the summation is also in the same interval. The minima of $E(\mathbf{s})$ now appear as maxima in $V(\mathbf{s})$. When all clauses are fully satisfied at a vertex, we have $V(\mathbf{s}) = M$, and the global maximum is attained here, i.e. $\mathbf{s} \in \mathcal{S}^*$.

The value of $V(\mathbf{s})$ at the points corresponding to binary solutions $\mathbf{s} \in \mathcal{V}$ indicates the cost for that binary input. A list of these costs could theoretically be expressed in a vector $\mathbf{c} \in \mathbb{R}^{2^n}$. The resulting vector $\mathbf{c}$ is loosely similar to a quantum state vector encoding the problem, in the sense that we could normalize and sample from it, and each entry could be mapped to the set of classical solutions. In this case, the modes of the resulting distribution would correspond to the optimal solutions, since the objective function attains maximum values at those points. It was shown in \cite{brassard2002quantum} that AA can be performed directly on the output state of an arbitrary quantum algorithm in order to boost the probabilities corresponding to valid solutions. The initial output state could be a superposition of all possible outcomes, such as with Grover's algorithm. The same concept turns out to apply to un-normalized vectors as well. In summary, if we had direct sample access to a vector like $\mathbf{c}$, constructed by sampling the value of the objective function at $2^n$ points, we could apply Grover-inspired oracle and diffusion operators to that vector to boost the amplitudes of entries corresponding to solutions only. In the classical setting we don't have access to such a vector, except through single queries to the objective function $V(\mathbf{s})$ for individual $\mathbf{s} \in \mathcal{V}$. But we can still define a set of transformations to the objective function such that each query for $\mathbf{s} \in \mathcal{V}$ will return a conditioned cost $V_{\ell}(\mathbf{s})$ according to an AA algorithm.
 \begin{algorithm}[H]
 \caption{AmplifySAT($\mathbf{s},  \ell$)}
 \begin{algorithmic}[1]
 \renewcommand{\algorithmicrequire}{\textbf{Input:}}
 \renewcommand{\algorithmicensure}{\textbf{Output:}}
 \REQUIRE $\mathbf{s} \in [-1, 1]^n$
 \ENSURE  $y \in \mathbb{R}$
 \STATE  $y \leftarrow f_0(\mathbf{s})$
  \IF {($\ell > 0$)}
  \FOR {$i = 0$ to $\ell$}
  \STATE $z \leftarrow 2\mathbb{E}_{\mathbf{s}}[T(\mathbf{s}) y] - T(\mathbf{s})y$
  \STATE $y \leftarrow z$ 
  \ENDFOR
  \ENDIF
 \RETURN $y$
 \end{algorithmic}
 \end{algorithm}

\subsection{Amplitude-Amplification-Inspired Conditioning}
The steps of AmplifySAT closely resemble its quantum analogs. In particular, we first propose multiplying $V(\mathbf{s})$ by a function $T(\mathbf{s})$ whose action on the vertices of the hypercube is identical to the action of a quantum oracle operator $U_{\omega}$ on a quantum state $|\psi\rangle \in \mathbb{C}^{2^n}$. We then follow this by a reflection about the average value of the function. A valid \textit{oracle function}, $T(\mathbf{s})$, must take the following values at the vertices $\mathbf{s} \in \mathcal{V}$: 
\begin{equation}
    \label{eq:oracle_cases}
    T(\mathbf{s}) = 
    \begin{cases}
    -1 \quad &\mathbf{s} \in \mathcal{S}^* \\
    1 \quad &\mathbf{s}   \in \mathcal{V} \setminus \mathcal{S}^* \, .
    \end{cases}
\end{equation}
The action of $T(\mathbf{s})$, then, when used to form $T(\mathbf{s})V(\mathbf{s})$ is to \textit{flip the sign} of the objective function at points where $\mathbf{s}$ corresponds to a satisfying classical input, i.e. $\mathbf{s} \in \mathcal{S}^*$, the set of satisfying inputs, and to do nothing to the function value at an unsatisfying argument. 

There are many ways to construct a function like $T(\mathbf{s})$. One such construction follows simply from the clause functions $K_m(\mathbf{s})$. In particular, we recognize that $K_{m}(\mathbf{s}) = 0$ for a fully satisfied clause and $K_{m}(\mathbf{s}) = 1$ for a fully unsatisfied clause. Then trivially we have the following relation for the product of clause functions:  
\begin{equation}
    \label{eq:prod_oracle}
    \prod_{m = 1}^{M} ( 1 - K_m(\mathbf{s})) = 
    \begin{cases}
    1  &\mathbf{s} \in \mathcal{S}^* \\ 
    0  &\mathbf{s} \in \mathcal{V} \setminus \mathcal{S}^*
    \end{cases}
    \, .
\end{equation}
The product of clause functions enforces the conjunction between clauses, since one or more unsatisfying clauses will render a $K_m(\mathbf{s})$ equal to $1$, and thus the entire product of Equation (\ref{eq:prod_oracle}) equal to $0$, or unsatisfied. From now on, we make the substitution $\tilde{K}_m(\mathbf{s}) = 1 - K_m(\mathbf{s})$. 
Equation (\ref{eq:prod_oracle}) can now be shifted and scaled to form the phase-flip-style oracle function $T(\mathbf{s})$ as
\begin{equation}
    \label{eq:oracle_fun}
    T(\mathbf{s}) = -2(\prod_{m = 1}^{M} \tilde{K}_m(\mathbf{s}) - \frac{1}{2})
\end{equation}
and $T(\mathbf{s})$ satisfies the description of Equation (\ref{eq:oracle_cases}). Of course, the oracle function is nonlinear in the variables, and may have many local extrema within the interior of the hypercube.

Other formulations of oracle functions are possible, so long as the result behaves like Equation (\ref{eq:oracle_cases}) at the vertices. Another satisfying choice is 
\begin{equation}
    T_2(\mathbf{s}) = -2(\min_{m \in [M] } \tilde{K}_m(\mathbf{s}) - \frac{1}{2} ) \, .
\end{equation}
Each unsatisfying solution $\mathbf{s} \in \mathcal{V} \setminus \mathcal{S}^* $ will have a fully unsatisfied clause, rendering $\tilde{K}_m(\mathbf{s}) = 0$ for some $m$. In contrast, at the satisfying solution we have $\min_{m \in [M] } \tilde{K}_m(\mathbf{s}) = 1$ and $T_1(\mathbf{s}^*) = -1$. This meets the criteria of Equation (\ref{eq:oracle_cases}). Another example that might be used for its gradient characteristics is 
\begin{equation}
    T_3(\mathbf{s}) = -2( \frac{V(\mathbf{s})}{M} \min_{m \in [M] } \tilde{K}_m(\mathbf{s}) - \frac{1}{2} ) \, .
\end{equation}
We expect that each choice of oracle function modifies the optimization landscape differently in the full algorithm,  affecting properties related to the gradient, smoothness, number of local minima, and thus the overall efficacy of solvers. 

After pointwise multiplication by $T(\mathbf{s})$, AmplifySAT continues mirroring AA algorithms by reflecting the function about its average value. But, before we detail this second stage, we also remark on the initial function by which to pointwise multiply $T(\mathbf{s})$. Throughout this section, we have suggested starting with objective function $V(\mathbf{s})$. But as previously mentioned, the effects of amplitude amplification essentially work on the output of any quantum algorithm $\mathcal{A}$ \cite{brassard2002quantum}. That is, an arbitrary quantum algorithm $\mathcal{A}$ producing a superposition of classical states $\mathcal{A}|0\rangle = \sum_i a_i |i\rangle $ can be subjected to AA to boost the probability of measuring a ``good" solution. In Grover's algorithm, $\mathcal{A} = \mathbf{H}_{2^n}$, with the interpretation that the Hadamard matrix is an algorithm producing a uniform distribution over all possible solutions.

In the analogous quantum-inspired setting, therefore, we might also consider different choices for the initial objective function, which we now refer to as $f_0(\mathbf{s})$, that reflect a state of prior knowledge. For example, the choice $f_0(\mathbf{s})  = 1$ represents an initial function that reflects no structure to the search problem and resulting optimization landscape, and is thus analogous to $\mathcal{A} = \mathbf{H}_{2^n}$. In this case, repeated application of the oracle function $T(\mathbf{s})$ and reflection about the average would do all the work of encoding information about the optimal solutions. We may alternatively proceed with our choice $f_0(\mathbf{s}) = V(\mathbf{s})$ as was suggested by the initial example, since it is an objective function that encodes useful initial information about the SAT problem; in particular, the cost of this function at each $\mathbf{s} \subseteq \mathcal{V}$ reports the number of clauses satisfied by that input. This introduces an intermediate measure of value of each solution, where points that have a higher value $V(\mathbf{s})$ are ``closer" to optimal." However, as we will see, it is also necessary in AmplifySAT to pre-compute the average value of the initial function $f_0(\mathbf{s})$, which can limit the available options. Fortunately, the choices $f_0(\mathbf{s}) = 1$ and $f_0(\mathbf{s}) = V(\mathbf{s})$ both admit pre-computed averages. 

\subsection{Averages}
In the previous discussion we see a melding of the ideas of a classical objective function and the $2^n$-ary outcomes that are encoded into a quantum state vector. The action of the oracle function $T(\mathbf{s})$ is identical to the action of the quantum oracle operator $U_{\omega}$ for a theoretical vector of costs. 
It turns out we can also apply a reflection about the average in our inspired algorithm as well. In fact, the average values are directly derivable for each iteration of the algorithm offline and in closed form, as long as the average values of $f_0(\mathbf{s})$ and $T(\mathbf{s})$ at the vertices are also known. For the basic construction of the algorithm, we continue to be primarily concerned with the behavior of the functions at the vertices only, because these are points corresponding to binary solutions. For this reason it is sufficient to define the average value of a function as  
\begin{equation}
    \label{eq:expec}
    \mathbb{E}_{\mathbf{s}}[f(\mathbf{s})] = \frac{1}{2^n} \sum_{\mathbf{s} \in \mathcal{V} } f(\mathbf{s}) \, .
\end{equation}
as opposed to a cumbersome integral over the entire interior $[-1, 1]^n$. We use the expectation operator for notational convenience; there are no random variables arising in Equation (\ref{eq:expec}). 

We may now define $s(\cdot) : \mathbb{R} \mapsto \mathbb{R}$ as the linear functional performing reflection about the average. That is, in general,
\begin{equation}
    s \circ f(\mathbf{s}) = 2\mathbb{E}_{\mathbf{s}}[ f(\mathbf{s})] - f(\mathbf{s}) \, . 
\end{equation}
One full iteration of AmplifySAT would thus produce the function
\begin{equation}
    \label{eq:first_averaging_op}
    s \circ (T(\mathbf{s}) f_0(\mathbf{s})) = 2 \mathbb{E}_{\mathbf{s}}[T(\mathbf{s}) f_0(\mathbf{s})] - T(\mathbf{s}) f_0(\mathbf{s})
\end{equation}
In a classical setting, we cannot apply such a reflection unless we know the value of the constant $\mathbb{E}_{\mathbf{s}}[T(\mathbf{s})f_0(\mathbf{s})]$ in advance, or we'd have to sample the function $T(\mathbf{s})f(\mathbf{s})$ to estimate the average value. Fortunately, as we now show, this average is available in closed form for the initial functions $f_0(\mathbf{s})$ of greatest interest, as long as the number of variables $n$, number of satisfying solutions $L$, number of clauses $M$, and averages $\mathbb{E}_{\mathbf{s}}[f_0(\mathbf{s})]$ and $\mathbb{E}_{\mathbf{s}}[T(\mathbf{s})]$ are known. Such prior assumptions are typical of AA algorithms and of satisfiability problems. Also, when the number of solutions $L$ is not known, there are still methods for applying AA algorithms effectively \cite{brassard2002quantum}. 

We can pre-compute $\mathbb{E}_{\mathbf{s}}[T(\mathbf{s})f_0(\mathbf{s})]$ for these problems by making the following observations. First, we note that the action of pointwise multiplying by $T(\mathbf{s})$ will flip the sign of the amplitude at the satisfying solutions. The value of $f_0(\mathbf{s})$ at these satisfying solutions is originally $f_0(\mathbf{s}^*)$. Thus the average value of the function with the flipped sign should change to
\begin{equation}
    \label{eq:exp_oracle}
    \mathbb{E}_{\mathbf{s}}[T(\mathbf{s}) f_0(\mathbf{s})] = \mathbb{E}_{\mathbf{s}}[f_0(\mathbf{s})] - 2 \frac{f_0(\mathbf{s}^*) L}{2^n} \, . 
\end{equation}
We subtract one term of $f_0(\mathbf{s}^*) L/2^n$ to remove the contributions of the positive-sign optimal terms in $\mathbb{E}_{\mathbf{s}}[f_0(\mathbf{s})]$, and then we subtract the second term to reflect that the sign at those points is now negative (i.e. amplitudes of $-1$). This gives us a closed form expression of $\mathbb{E}_{\mathbf{s}}[T(\mathbf{s}) f_0(\mathbf{s})]$ for many choices of $f_0(\mathbf{s})$. For instance, with our choice of $f_0(\mathbf{s}) = 1$ we have $\mathbb{E}_{\mathbf{s}}[f_0(\mathbf{s})] = 1$ and $f_0(\mathbf{s}^*) = 1$. 

For the case $f_0(\mathbf{s}) = V(\mathbf{s})$ we can easily deduce that $f_0(\mathbf{s}^*) = M$ and also calculate that 
\begin{subequations}
    \begin{alignat}{2}
        \mathbb{E}_{\mathbf{s}}[V(\mathbf{s})] &= \mathbb{E}_{\mathbf{s}}\biggl[ \sum_{m = 1}^{M} \bigl(1 -  2^{-k}\prod_{i = 1}^{n} (1 - c_{m,i}s_i) \bigr)  \biggr] \\ 
        &= \sum_{m = 1}^{M} \mathbb{E}_{\mathbf{s}}\biggl[ \bigl( 1 -  2^{-k}\prod_{i = 1}^{n} (1 - c_{m,i}s_i)\bigr) \biggr] \\ 
        &= M - \sum_{m = 1}^{M} 2^{-k}\mathbb{E}_{\mathbf{s}}[\prod_{i = 1}^{n} (1 - c_{m , i}s_i)] \, . 
    \end{alignat}
\end{subequations}
Recall that the product $K_m(\mathbf{s}) = 2^{-k}\prod_{i = 1}^{n} (1 - c_{m , i}s_i)$ equals zero when the clause is satisfied and $1$ when unsatisfied. Any $k$ variables or their negations featured in a clause will be unsatisfied for exactly $2^{n - k}$ cases, since for any disjunction of $k$ variables, there is only one combination of those variables that makes that disjunction equal to $0$. That leaves $2^n - 2^{n - k}$ cases for which the clause is satisfied. In expectation, therefore, we have for any clause function that 
\begin{subequations}
    \begin{alignat}{2}
        \mathbb{E}_{\mathbf{s}}[2^{-k} \prod_{i = 1}^{n} (1 - c_{m , i}s_i)] &= 0 \cdot \frac{2^n - 2^{n - k}}{2^n} + 1\cdot \frac{2^{n - k}}{2^n}  \\
        &= 2^{-k} \,, 
    \end{alignat}    
\end{subequations}
which gives the final expression
\begin{equation}
    \label{eq:E_Vs}
    \mathbb{E}_{\mathbf{s}}[V(\mathbf{s})] = M(1 - 2^{-k}) \, . 
\end{equation}
Substituting these expressions into Equation (\ref{eq:exp_oracle}) gives 
\begin{equation}
    \mathbb{E}_{\mathbf{s}}[T(\mathbf{s}) V(\mathbf{s})] = M(1 - 2^{-k}) - 2\frac{ML}{2^n} \, .
\end{equation}

This process needs to be iterated for the full AA effect. Fortunately, the iterations continue to admit a closed form expression with no additional assumptions. It can be shown that for an initial objective function $f_0(\mathbf{s})$ the following relationships hold. First we define the evolved function after each application of the pair of oracle and diffusion operations: 
\begin{equation}
    \label{eq:recursive_cost}
    f_{\ell}(\mathbf{s}) = 2\mathbb{E}_{\mathbf{s}}[T(\mathbf{s})f_{\ell - 1}(\mathbf{s})] - T(\mathbf{s})f_{\ell - 1}(\mathbf{s}) \, .
\end{equation}
Letting $W_{\ell}(\mathbf{s}) = T(\mathbf{s})f_{\ell}(\mathbf{s})$ we have, equivalently,
\begin{equation}
    \label{eq:cond_fn}
    f_{\ell}(\mathbf{s}) = 2\mathbb{E}_{\mathbf{s}}[W_{\ell-1}(\mathbf{s})] - W_{\ell - 1}(\mathbf{s}) \, . 
\end{equation}
The expression for $\mathbb{E}_{\mathbf{s}}[W_0]$ is given in Equation (\ref{eq:exp_oracle}). In general, we have 
\begin{equation}
    \mathbb{E}[W_{\ell}(\mathbf{s})] = 2\mathbb{E}_{\mathbf{s}}[W_{\ell - 1}] \mathbb{E}_{\mathbf{s}}[T(\mathbf{s})] - \mathbb{E}_{\mathbf{s}}[T(\mathbf{s}) W_{\ell - 1}] \, , 
\end{equation}
or, equivalently, for $\ell \geq 2$,
\begin{equation}
    \label{eq:exp_recur}
    \mathbb{E}_{\mathbf{s}}[W_{\ell}(\mathbf{s})] = 2\mathbb{E}[W_{\ell - 1}(\mathbf{s})] \mathbb{E}_{\mathbf{s}}[T(\mathbf{s})] - \mathbb{E}_{\mathbf{s}}[W_{\ell - 2}(\mathbf{s})]
\end{equation}
Equation (\ref{eq:exp_recur}) is obtained by expanding $ \mathbb{E}[W_{\ell}(\mathbf{s})]$ using Equation (\ref{eq:recursive_cost}) and applying linearity of expectation, that is, 
\begin{equation}
    \begin{split}
    \mathbb{E}_{\mathbf{s}}\biggl[T(\mathbf{s})&\bigr[2\mathbb{E}_{\mathbf{s}}[W_{\ell-1}(\mathbf{s})] - W_{\ell - 1}(\mathbf{s})\bigl]\biggr] \\ 
    &= 2\mathbb{E}_{\mathbf{s}}[W_{\ell-1}(\mathbf{s})]\mathbb{E}[T(\mathbf{s})] - \mathbb{E}_{\mathbf{s}}[T(\mathbf{s})W_{\ell - 1}(\mathbf{s})] \, . 
    \end{split}
\end{equation}
We then use the fact that 
\begin{alignat}{2}
    \mathbb{E}_{\mathbf{s}}[T(\mathbf{s}) W_{\ell - 1}(\mathbf{s})] &= \mathbb{E}_{\mathbf{s}}[T(\mathbf{s})^2 [2\mathbb{E}_{\mathbf{s}}[W_{\ell - 2}] - W_{\ell - 2}]] \\ 
    &= \mathbb{E}_{\mathbf{s}}[W_{\ell - 2}(\mathbf{s})]
\end{alignat}
since $T(\mathbf{s})^2 = 1$ at any vertex. This gives a method for computing all of the required expectations of the form $\mathbb{E}_{\mathbf{s}}[W_{\ell}(\mathbf{s})]$ in advance using only knowledge of $L, n$, and $M$, and known averages. The expectations can then be substituted as required in computations of the iterates in Equation (\ref{eq:cond_fn}).

\section{Conclusion}
\label{sec:conclusion}

\begin{figure} 
    \centering
  \subfloat[$f_0(\mathbf{s})$ at vertices \label{1a}]{%
       \includegraphics[width=0.5\linewidth]{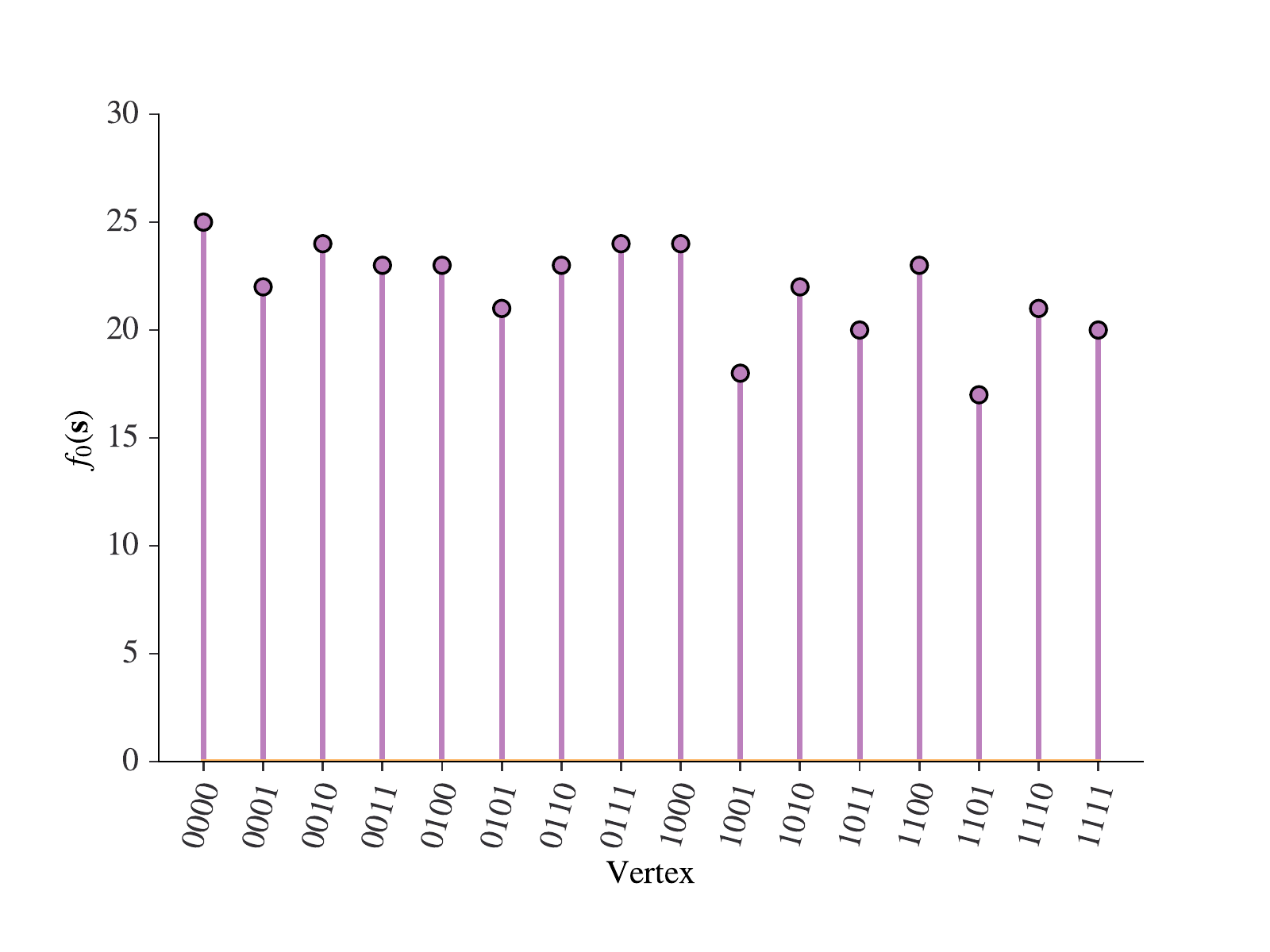}}
  \subfloat[Costs after 1 iteration \label{1b}]{%
        \includegraphics[width=0.5\linewidth]{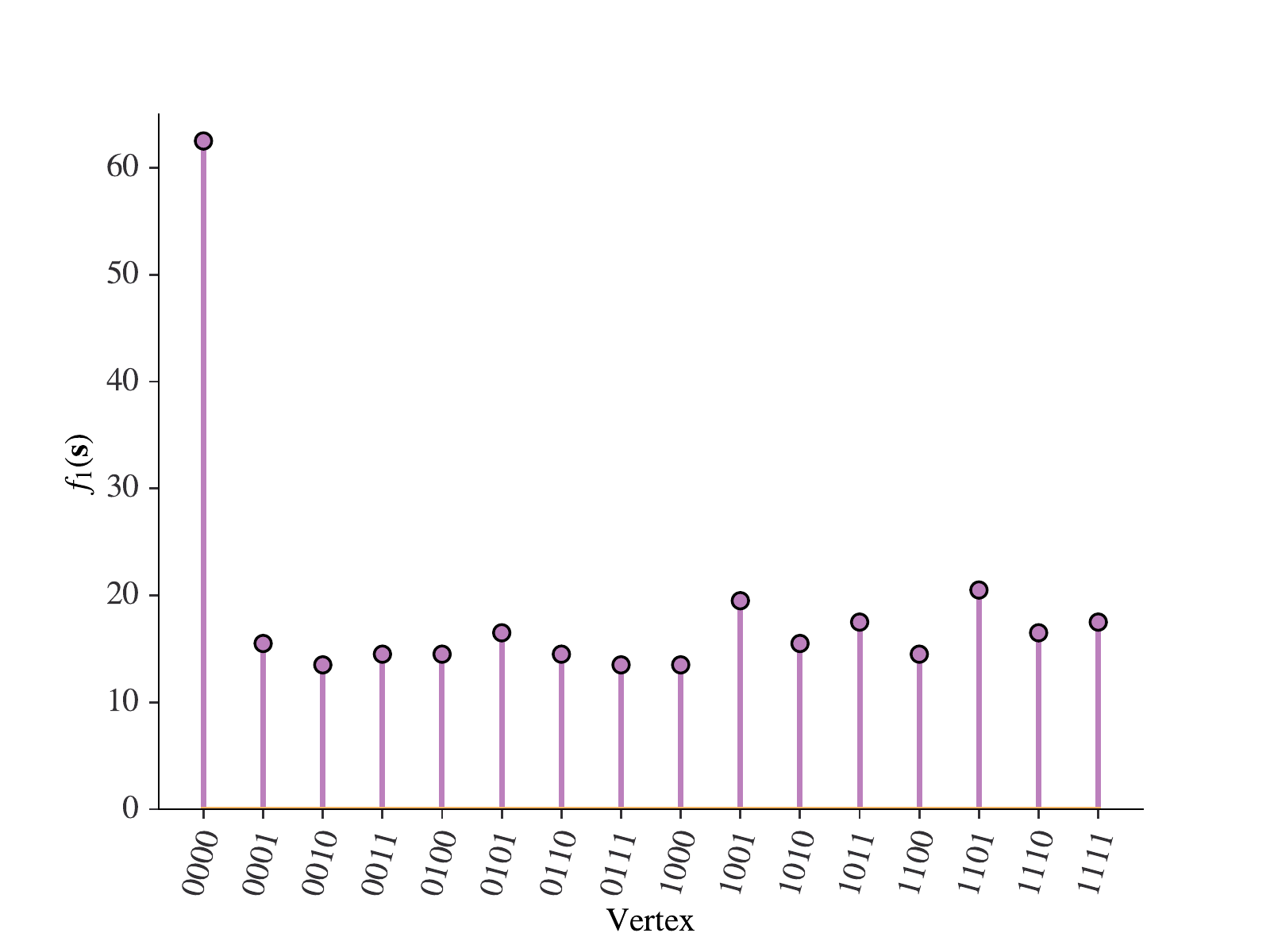}}
  \\
  \subfloat[Costs after 2 iterations \label{1c}]{%
        \includegraphics[width=0.5\linewidth]{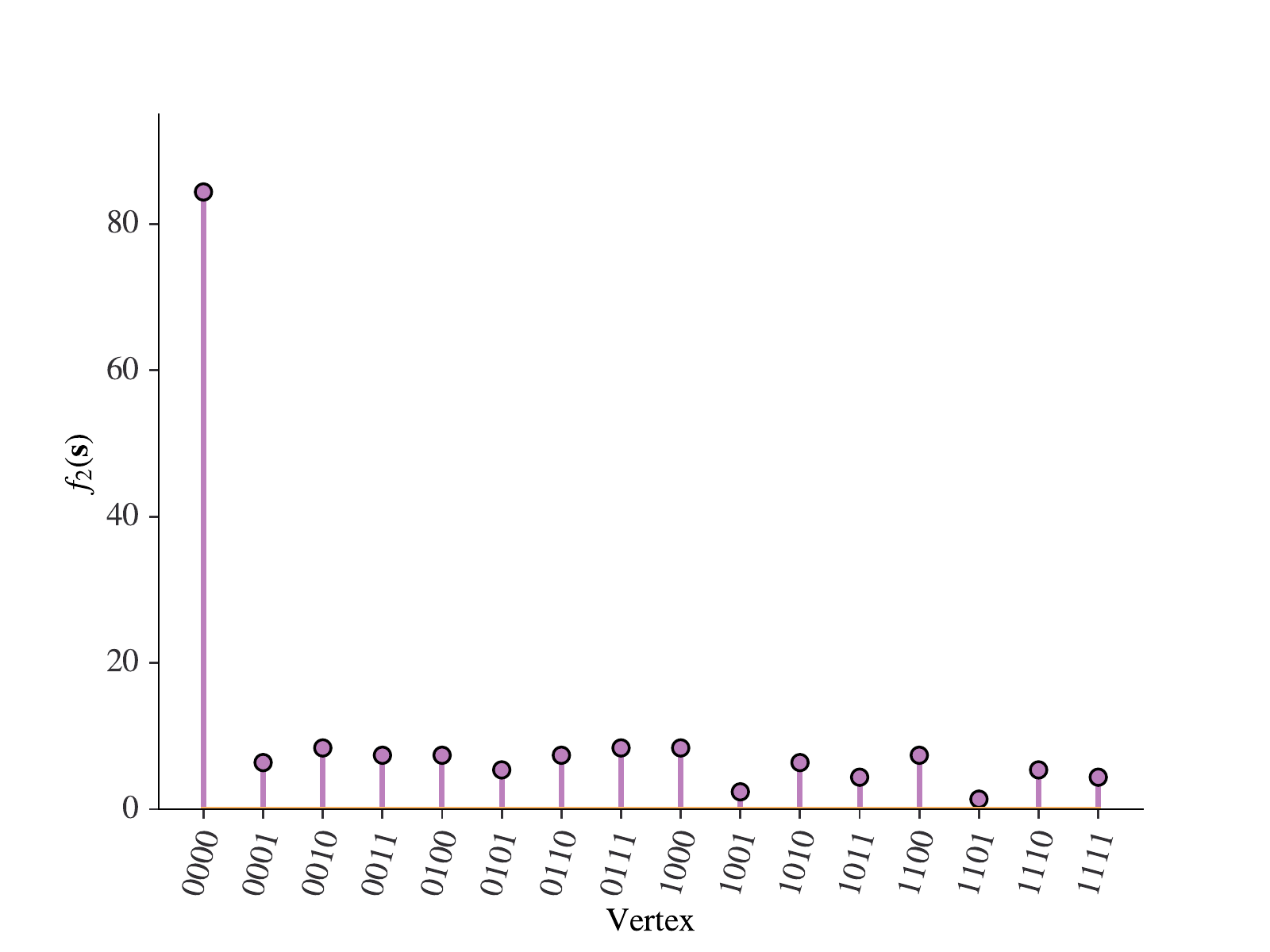}}
  \subfloat[Costs after 3 iterations \label{1d}]{%
        \includegraphics[width=0.5\linewidth]{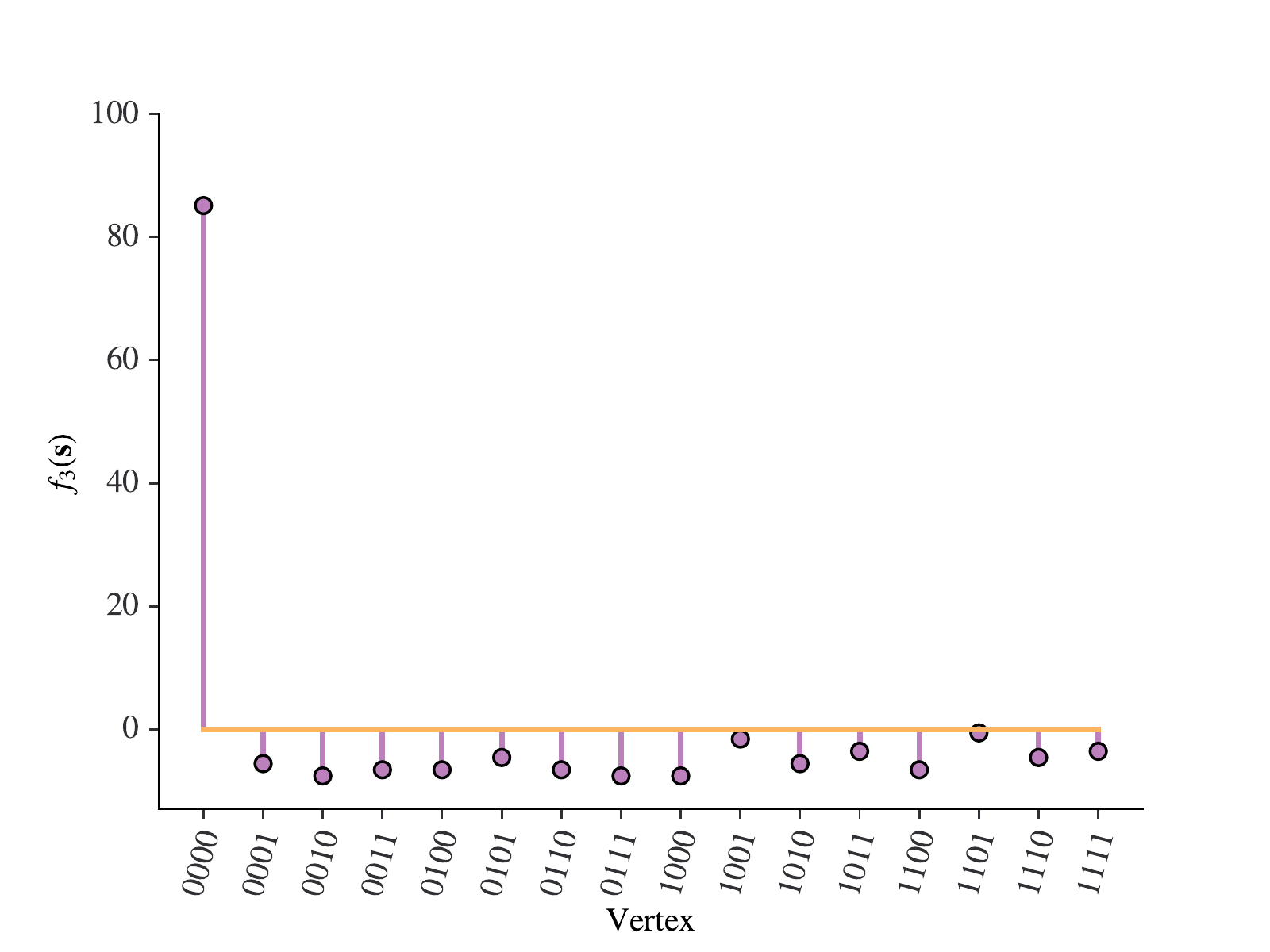}}
  \caption{\label{fig:costs} (color online) Graphical depiction of costs for an example $k$-SAT problem. The costs are evaluated at each of the vertices in $\mathcal{V}$, which correspond to the costs at binary solutions, and displayed sequentially. As such they are labeled with their binary equivalent representation. For this example, we generated a $3$-SAT problem with $n = 4$ variables, $L = 1$ solution, and $M = 25$ clauses. We used the objective function formulation $V(\mathbf{s})$ to serve as $f_0(\mathbf{s})$, and we used the oracle function of Equation (\ref{eq:oracle_fun}) to evaluate the iterates.}
  \label{fig1} 
\end{figure}

Figure \ref{fig:costs} demonstrates the transformed costs at the vertices that result from applying  conditioning to an example problem. The number of iterations that provides the most amplification is consistent with theoretical results about AA, i.e. $\ell^* = \floor{\pi/4 \sqrt{2^n/L}}$ iterations \cite{brassard2002quantum}. For this problem, the number of iterations to apply for maximum amplification was $3$. The most prominent effect of the conditioning is to greatly increase the cost at $f(\mathbf{s}^*)$ relative to the cost at all non-satisfying vertices. The effects have discernible impact on a variety of parameters relevant to non-convex solvers, but incur the processing of applying AmplifySAT, for which the optimal number of iterations appears to grow as $\mathcal{O}(\sqrt{N})$. Although, as Figure \ref{1b} and \ref{1c} highlight, there is still a significant amplification achievable when performing fewer than the ideal number of iterations, and the potential for a tradeoff emerges between the cost of conditioning and the degree of amplification desired. 

In simulated annealing, the \textit{maximum depth} of any local minimum determines the cooling schedule of the temperature parameter, and also impacts the convergence rate \cite{hajek1988cooling, mitra1986convergence}. It appears that an energy landscape can be engineered from $f_{\ell^*}(\mathbf{s})$ with local minima of reduced depth, arising directly from the effects of the conditioning and potentially yielding a faster convergence rate to the global optimum when using simulated annealing. Similarly, solvers based on the gradient of the objective function would use $\nabla f_{\ell^*}(\mathbf{s})$ after applying the algorithm, as opposed to $\nabla f_0(\mathbf{s})$. A more detailed analysis of how the gradient evolves with AmplifySAT will help illuminate whether the conditioned gradient improves convergence properties to the global minimum. It is not clear whether the conditioning would improve the performance of these solvers, as the choice of oracle function $T(\mathbf{s})$ could introduce additional local minima on the interior, or create other undesirable features across the optimization landscape. In this sense an informed choice of oracle function $T(\mathbf{s})$ could also greatly determine the usefulness of AmplifySAT with gradient-based solvers. 

Also of interest is whether an AmplifySAT-conditioned function can be constructed as a continuous-time dynamical system, as was achieved with the original UniSAT objective function \cite{ercsey2011optimization, yin2017efficient}. A successful conversion would require another reformulation of the objective function, as the dynamical systems are energy-minimizing (dissipative), while much of this work has focused on amplifying, that is, increasing global maxima. However, if it is possible to express the objective function as an energy-minimization problem, while also improving its energy landscape through AmplifySAT, then this could improve the performance of such solvers. The iterative nature of AmplifySAT raises the possibility of a tradeoff between conditioning steps of the objective function and the available energy budget.

In summary, a more comprehensive analysis will be useful to analyze the performance of various non-convex solvers when paired with AmplifySAT. A numerical investigation of the performance of solvers using AmplifySAT would help delineate any performance differences and would inform theoretical developments. While there is not currently a direct theoretical relationship between the conditioning applied by AmplifySAT and the performance of these solvers, it appears that the algorithm directly modifies parameters of the objective function that affect convergence to the global optima. The performance improvement, if it exists, will heavily depend on design decisions regarding the original objective function, oracle function, the number of iterations to apply, and the non-convex solution method pursued.


\section*{Acknowledgment}
The authors would like to thank The Office of Naval Research for their support of this work under Grant No.\ N00014-17-1-2107, as well as Applied Research Laboratories, The University of Texas at Austin for an internal research and development grant. 


\bibliographystyle{IEEEtran}
\bibliography{grover_inspired}
\nocite{o2014analysis}

\end{document}